\documentclass[twocolumn]{aastex631}
\usepackage{
    amsmath,
    amssymb,
    amsfonts,
    graphicx,
    hyperref,
    xcolor,
    soul,
    multirow,
    bm,
    verbatim
}
\newcommand{\ls}{\textcolor{red}}

\definecolor{haykcolor}{HTML}{bf2844}

\graphicspath{{./figures/}}

\begin{document}

\title{Kinetic simulations of the Kruskal-Schwarzchild instability in accelerating striped outflows \\ I: Dynamics and energy dissipation}

\correspondingauthor{William Groger}
\email{weg2114@columbia.edu}

\author[0009-0009-0500-4514]{William Groger}
\affiliation{Department of Astronomy, Columbia University, New York, NY, 10027, USA}
\affiliation{Columbia Astrophysics Laboratory, Columbia University, New York, NY 10027, USA}

\author[0000-0001-8939-6862]{Hayk Hakobyan}
\affiliation{Columbia Astrophysics Laboratory, Columbia University, New York, NY 10027, USA}
\affiliation{Computational Sciences Department, Princeton Plasma Physics Laboratory (PPPL), Princeton, NJ 08540, USA}
\author[0000-0002-1227-2754]{Lorenzo Sironi}
\affiliation{Department of Astronomy, Columbia University, New York, NY, 10027, USA}
\affiliation{Columbia Astrophysics Laboratory, Columbia University, New York, NY 10027, USA}

\affiliation{Center for Computational Astrophysics, Flatiron Institute, 162 5th Avenue, New York, NY 10010, USA}

\begin{abstract}
Astrophysical relativistic outflows are launched as Poynting-flux-dominated, yet the mechanism governing efficient magnetic dissipation, which powers the observed emission, is still poorly understood. We study magnetic energy dissipation in relativistic ``striped'' jets, which host current sheets separating magnetically dominated regions with opposite field polarity. The effective gravity force $g$ in the rest frame of accelerating jets drives the Kruskal-Schwarzschild instability (KSI), a magnetic analogue of the Rayleigh-Taylor instability. By means of 2D and 3D particle-in-cell simulations, we study the linear and non-linear evolution of the KSI. The linear stage is well described by linear stability analysis. The non-linear stages of the KSI generate thin (skin-depth-thick) current layers, with length comparable to the dominant KSI wavelength. There, the relativistic drift-kink mode and the tearing mode drive efficient magnetic dissipation. The dissipation rate can be cast as an increase in the effective width $\Delta_{\rm eff}$ of the dissipative region, which follows $d\Delta_{\rm eff}/dt\simeq 0.05 \sqrt{\Delta_{\rm eff}\,g}$. Our results have important implications for the location of the dissipation region in gamma-ray burst and AGN jets.
\end{abstract}

\keywords{
\href{http://astrothesaurus.org/uat/739}{High energy astrophysics (739)}; 
\href{http://astrothesaurus.org/uat/1261}{Plasma astrophysics (1261)}; 
\href{http://astrothesaurus.org/uat/288}{Compact objects (288)}; 
\href{http://astrothesaurus.org/uat/994}{Magnetic fields (994)}
}

\section{Introduction}
\label{sec:introduction}
Astrophysical relativistic outflows, commonly observed in pulsar winds, gamma-ray bursts (GRBs) and active galactic nuclei (AGN), are launched as Poynting-flux dominated, driven by a strong magnetic field threading a rotating compact object---either a spinning black hole and/or its accretion disk \citep{blandford_77,blandford_payne_82,meier_01,vlahakis_03} or a magnetized neutron star \citep{rees_gunn_74,kennel_coroniti_84,bogovalov_99,usov_92,metzger_11}. The flow starts as magnetically dominated, and some form of internal dissipation of the dominant magnetic energy is required to mediate the powerful, rapid release of energy inferred from observations.

Magnetic dissipation is more likely to be triggered when small-scale structures with oppositely-directed fields preexist in the flow. In pulsar winds, such a structure arises naturally because the magnetic field near the equatorial plane changes sign every half of the pulsar period, creating the so-called \textit{striped pulsar wind} of magnetically-dominated regions separated by thin current sheets carrying hot pair-plasma. It is the annihilation of these oppositely directed fields that provides the main energy conversion mechanisms in pulsar winds. Magnetic dissipation of the stripes starts close to the light cylinder \citep{lyubarsky_kirk_01, kirk_03, petri_05}, 
potentially exhausting most of the carried Poynting flux well before the pulsar wind termination shock
\citep{cerutti_philippov_17, Cerutti_20, hakobyan_spitkovsky_23}.

Outflows with alternating fields could also arise in accreting systems -- GRBs and AGN jets -- if the magnetic field in the central engine changes sign \citep{drenkhahn_02a,drenkhahn_02b,giannios_uzdensky_19,zhang_giannios_21,2021MNRAS.508.1241C}. 
Efficient dissipation of the alternating fields in such \textit{striped jets} would occur when the drift velocity of the current carriers approaches the speed of light (the so-called ``charge starvation'' regime), or equivalently when the particle Larmor radius is comparable to the thickness of the sheet. In GRBs and AGNs, the relativistic jet is expected to be heavily loaded with the plasma from the accretion disk (and in GRBs, also from the progenitor star \citep{levinson_eichler_03}), so charge starvation is achieved only at extremely large distances. Therefore, the onset of magnetic dissipation in GRB and AGN jets occurs too far from the central engine to explain the observed high-energy emission.

This motivated \citet{Lyubarsky2010} to suggest that magnetic dissipation in striped relativistic jets could be facilitated by the Kruskal-Schwarzschild instability (KSI), an analogue of the Rayleigh-Taylor instability (RTI) in strongly magnetized flows. It was shown that as the flow accelerates, the current layer in its comoving frame experiences an effective gravitational acceleration $g=c^2 d\Gamma/dr$ in the opposite direction (here, $\Gamma$ is the bulk Lorentz factor of the jet, and $r$ is the distance from the central engine). Since the enthalpy density of the relativistically hot plasma in the current layer is larger than that of the cold magnetized plasma below it, the current sheet becomes susceptible to the KSI just like the interface between a lighter fluid below a heavier one would be to the RTI. As the plasma drips out of the layer in-between the magnetic field lines, it intermixes regions of opposing field polarities driving magnetic energy dissipation,
which in turn leads to further acceleration of the flow, so the process is self-sustaining  \citep{Lyubarsky2010}.

The structure and temporal evolution of the KSI was studied using 2D relativistic magnetohydrodynamic (MHD) simulations by \citet{Gill2018}, who confirmed that the instability growth rate  matches the predictions from the linear stability analysis. They also measured the magnetic dissipation rate, finding that it corresponds to an effective bulk velocity inflow into the dissipation region of $\lesssim 0.005 \, c$ --- too slow to explain efficient dissipation in GRB and AGN jets.
In this work, we study for the first time the linear and non-linear evolution of the KSI using fully-kinetic particle-in-cell (PIC) simulations. As compared to MHD, a kinetic approach has several advantages: (\textit{i}) it captures the development of kinetic instabilities that cannot be described in MHD, e.g., the relativistic drift-kink instability \citep{zenitani_07}; (\textit{ii}) it allows to properly model the physics of collisionless reconnection---in fact, resistive MHD approaches yield reconnection rates that are an order of magnitude lower than equivalent kinetic calculations \citep{birn_01, cassak_review}; (\textit{iii}) it can naturally describe the formation of non-thermal particle distributions. 
 
The paper is organized as follows. In Section \ref{sec:setup} we present the simulation setup, introducing the relevant parameters of the problem. In Section \ref{sec:analytic} we extend the linear analysis of \cite{Lyubarsky2010}---which was tailored to a relativistically hot layer---to a more general case. In Section \ref{sec:results} we present results from our 2D and 3D simulations, with particular focus on the non-linear dynamics and the efficiency of magnetic dissipation. Section \ref{sec:conclusion} summarizes our main findings and discusses their implications for GRB and AGN jets.

\section{Simulation setup}
\label{sec:setup}
We perform 2D and 3D simulations using the \texttt{Tristan v2} particle-in-cell (PIC) code \citep{tristan_v2}. In both 2D and 3D, our Cartesian domain is periodic in all directions. The 2D domain has dimensions $L_x\times L_y=5600\times10000$ cells in $x$ and $y$, respectively (with larger 2D runs having dimensions of $L_x\times L_y =8400\times 15000$). In 3D, our simulation has dimensions of $L_x\times L_y\times L_z = 1400\times2500\times1440$ cells. The domains are initialized with a magnetic field: 
\begin{equation*}
    \bm{B} = \hat{\bm{z}}\,B_0 \left(-\tanh{\left[\frac{y-y_{\mathrm{cs},1}}{\Delta_{\rm cs}}\right]}+\tanh{\left[\frac{y-y_{\mathrm{cs},2}}{\Delta_{\rm cs}}\right]}+1\right),
\end{equation*}

\noindent the direction of which switches at specific locations $y=\{y_{\mathrm{cs},1},y_{\mathrm{cs},2}\}$ over a characteristic width of $\Delta_{\rm cs}$. The background is filled with cold ($T_0 \ll m_e c^2$)\footnote{We use $k_B=1$, quoting all temperatures in units of energy.} uniform electron-positron plasma of number density $n_0$. The magnetization of the background, $\sigma_0\equiv B_0^2 / (4\pi n_0 m_e c^2)$, is fixed at $\sigma_0=10$ for all of our simulations,  in both 2D and 3D. The skin-depth of the background plasma, $d_0\equiv \left(m_e c^2/(4\pi n_0 e^2)\right)^{1/2}$, is typically resolved with $5$ grid cells, $\Delta x$, in 2D (larger runs have $d_0 = 2.5\,\Delta x$), and $2.5$ grid cells in 3D. Unless otherwise specified, we will report all the lengthscales in units of $d_0$. In 2D, the background number density $n_0$ is sampled with $8$ particles per cell, and with $4$ in 3D.


\begin{figure}[htb!]
    \centering
    \includegraphics[width=\columnwidth]{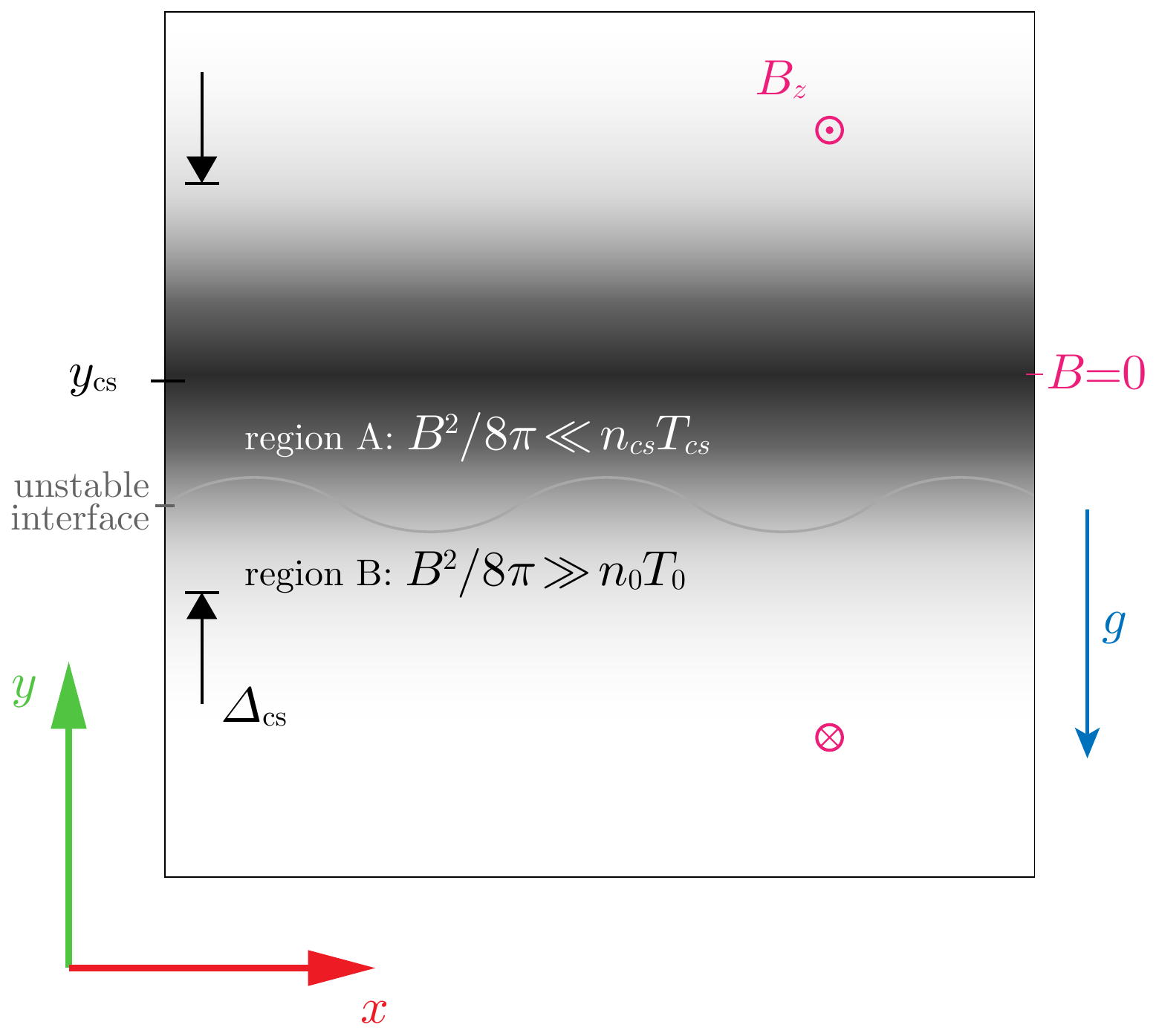}
    \caption{Schematic illustration of the simulation setup. Close to the middle of the current layer, where the magnetic field vanishes, the enthalpy is dominated by plasma pressure (region \textbf{A}), while below that region it is dominated by magnetic pressure (region \textbf{B}). The unstable interface lies between the two regions.}
    \label{fig:schematics}
\end{figure}

On top of the background plasma, we also initialize two overdense layers in the $x$-$z$ plane, having width $\Delta_{\rm cs}$ and the following density profile: $n = n_{\rm cs}/\cosh^2{\left[(y - y_{\mathrm{cs},i})/\Delta_{\rm cs}\right]}$ with $i=\{1,2\}$, where $n_{\rm cs}/n_0$ is the current layer overdensity w.r.t. the background plasma. The overdensity is set to $n_{\rm cs}/n_0=12$ in most of our runs. The overdense plasma in the layers also provides net current density that supports the magnetic field polarity switch across the sheets. To compensate the magnetic pressure outside the layer, the overdense population in the layer has a temperature of $T_{\rm cs} = B_0^2/(8\pi n_{\rm cs})$. We present simulations with $n_{\rm cs}/n_0=3$ and $12$ (the former case is described in  Appendix~\ref{sec:low-overdensity}), however we varied this parameter from $1.5$ to $30$ in simulations not presented in this paper, to validate the temperature dependence of the KSI growth rate.  We use $\Delta_0 = 20\,d_0$ as the fiducial current sheet width $\Delta_{\rm cs}$. In the 2D runs we set $\Delta_{\rm cs} / \Delta_0$ to $1$, $2$, or $4$.


The configuration described above is an exact kinetic equilibrium (double Harris layer), provided that $|y_{\mathrm{cs},1}-y_{\mathrm{cs},2}|\gg \Delta_{\rm cs} \gg d_{\rm cs}$, where $d_{\rm cs}\equiv \left(m_e c^2/(4\pi n_{\rm cs} e^2)\right)^{1/2}$ is the plasma skin-depth in the current layers. We also add a constant gravity field, acting on every simulation particle in the domain: $\bm{g} =\pm g\,\hat{\bm{y}}$, with the sign chosen in such a way that the force is directed towards the $y=L_y/2$ plane \citep{zhdankin_23}. All our results are derived using the upper half of the domain. The gravitational free fall acceleration, $g$, is parametrized similar to the convention used for radiative drag forces \citep[see, e.g.,][]{2014ApJ...780....3U}, where we set the fiducial value of $g_B$ by equating the force imposed by an electric field of magnitude $\eta_{\rm rec} B_0$ to the gravity force. Thus, we define the fiducial acceleration, $g_B \equiv \eta_{\rm rec} \,c \,\omega_{B,0}$, where $\omega_{B,0} \equiv |e|B_0/m_e c$, and $\eta_{\rm rec}=0.1$ is the fiducial reconnection rate. By further defining the value $g_0 = g_B/400$ (to make gravity much weaker than any electromagnetic forces), we parameterize the gravitational acceleration with a dimensionless factor: $g/g_0$. In the 2D runs we set this value to $1$, $1/2$, or $1/4$, while in 3D we pick $g/g_0=1$. We work in the weak gravity regime, studied analytically by \cite{Lyubarsky2010}, where $g\ll c^2/\Delta_{\rm cs}$; as an example, for $\Delta_{\rm cs}/\Delta_0=4$, which is the largest current sheet width we have considered, this inequality translates to $g/g_0\ll 16$, meaning that all the simulations presented in this work satisfy this inequality. To avoid artificial transients due to abruptly applying gravity to an otherwise stable equilibrium, we start all of our simulations with $g(t=0)=0$, gradually turning it on to its maximum value as $g(t) = g\left[1-\cos(\pi t/t_g)\right]/2$, for $t<t_g$, and $g(t) = g$, otherwise. The value of $t_g = L_x/c$ is short enough that the dynamics at early times does not affect the long-term evolution of the system, which we follow until  $t\gg t_g$.

Table~\ref{table:params} lists all the numerical and physical parameters of our simulations. Triple-dots indicate that we varied that  parameter in the specified range either for convergence study, or while exploring the parameter space.

\begin{table}[h!]
\begin{tabular}{ |l||c|c| } 
\hline
 parameter [units] & 2D & 3D \\
\hline \hline
box size [$d_0$] & $(1120, 2000)$, & $(560,1000,576)$ \\
& $(3360, 6000)$ & \\
\hline
particles per cell & $8$ & $4$ \\ 
\hline
$c\Delta t$ [$\Delta x$] & \multicolumn{2}{c|}{$0.45$} \\ 
\hline
$T_0$ [$m_e c^2$] & \multicolumn{2}{c|}{$10^{-4}$} \\ 
\hline
$\sigma_0$ & \multicolumn{2}{c|}{$10$} \\ 
\hline
$d_0$ [$\Delta x$] & $2.5...5$ & $2.5$ \\ 
\hline
$y_{\mathrm{cs},1}$, $y_{\mathrm{cs},2}$ [$L_y$] & (0.25, 0.75) & (0.15, 0.85) \\ 
\hline
$n_{\rm cs}/n_0$ & $3...12$ & $12$ \\ 
\hline
$\Delta_{\rm cs}$ [$d_0$] & $20...80$ & $40$ \\ 
\hline
$g/g_0$  & $0.25...1$ & $1$ \\ 
\hline
\end{tabular}
\caption{Numerical values for all the parameters used in our 2D and 3D simulations.}
\label{table:params}
\end{table}

\section{Analytical growth rate}
\label{sec:analytic}
To estimate the  growth rate of the gravity-driven KSI, we follow the approach by \cite{Lyubarsky2010}, where the assumption of weak gravity is employed: $g\Delta_{\rm cs}/c^2\ll 1$. In this case, the dispersion relation for the growth of a $k_x=k$ mode can be written as:

\begin{equation}
    \frac{\omega^4}{k^2g^2} = \frac{1-e^{-2k\Delta_{\rm cs}}}{ \tilde{h}^2-e^{-2k\Delta_{\rm cs}}}\approx 
    \begin{cases}
        \frac{1}{\tilde{h}^2},~~~&\textrm{if}~k\Delta_{\rm cs}\gg 1;\\
        \frac{2k\Delta_{\rm cs}}{\tilde{h}^2-1},~~~&\textrm{if}~k\Delta_{\rm cs}\ll 1,
    \end{cases}
    \label{eq:KS_dispersion}
\end{equation}
where $ \tilde{h} \equiv (h_A+h_B)/(h_A-h_B)$, and $h_A$ and $h_B$ are the enthalpy densities of the two regions on either side of the instability interface. These are shown in Fig.~\ref{fig:schematics} as \textbf{A}, the hot unmagnetized region dominated by plasma pressure, and \textbf{B}, the cold magnetized region dominated by magnetic pressure. As the dispersion relation in Eq.~\eqref{eq:KS_dispersion} always admits an unstable mode, the growth rate of the instability can be found as $\eta_g = -\mathbb{I}\mathrm{m}\left\{\omega\right\}$. In general, $h_I = n_Im_e c^2+\Gamma_I p_I/(\Gamma_I - 1) $ with $I=\{A,B\}$, where $\Gamma_I$ is the effective adiabatic index in the given region, while $n_I$ and $p_I = n_I T_I+B_I^2/8\pi$ are the plasma number density and the total pressure, respectively. 

\begin{figure}
    \centering
    \includegraphics[width=\columnwidth]{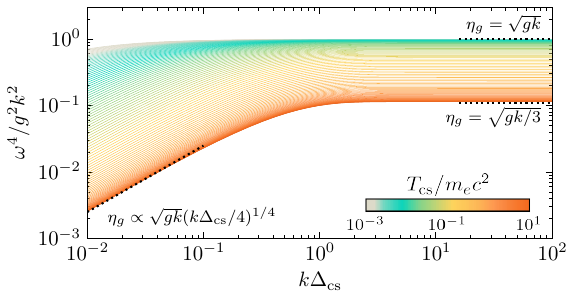}
    \caption{Dispersion relation of the KSI for different values of the temperature in region \textbf{A} (i.e., in the current sheet). Asymptotic predictions are shown with dotted black lines. For the purposes of this plot, we interpolate $\Gamma_A\approx 5/3$ to $\Gamma_A\approx 4/3$ for values of $\Theta_{\rm cs}\ll 1$ to $\Theta_{\rm cs}\gg 1$ as $\Gamma_A = (4/3 - 5/3) (1 + \tanh\{\log_{10}{\Theta_{\rm cs}} + 1\}) / 2 + 5/3$.}
    \label{fig:dispersion}
\end{figure}

Region \textbf{B} is mostly filled with the background plasma, $n_B\approx n_0$ and $T_B\approx T_0$, which is cold and highly magnetized: $n_0 T_0 \ll n_0 m_e c^2\ll B_0^2/8\pi$. Thus, the main contributor to the pressure in that region is the magnetic field: $h_B\approx 2\,p_B\approx B_0^2/4\pi$, where we used an adiabatic index of $\Gamma_B\approx 2$.
In region \textbf{A}, on the other hand, the magnetic field is subdominant, while the pressure is provided primarily by the current-layer particles, $n_A\approx n_{\rm cs}$, with a temperature of $T_A\approx T_{\rm cs} = (1/2)\,\sigma_0 m_e c^2\, (n_0 / n_{\rm cs})$ (which could be sub-relativistic or ultra-relativistic). Thus, defining $\tilde{\Gamma}_{\rm cs}\equiv \Gamma_A/(\Gamma_A-1)$ for the plasma in the layer, we can rewrite the expression for $\tilde{h}$ in the following form:
\begin{equation}
    |\tilde{h}| \approx \left|\frac{1+\Theta_{\rm cs}(\tilde{\Gamma}_{\rm cs}+2)}{1+ \Theta_{\rm cs}(\tilde{\Gamma}_{\rm cs}-2)}\right|\approx
    \begin{cases}
        1,~~~&\textrm{if}~\Theta_{\rm cs}\ll1;\\
        3,~~~&\textrm{if}~\Theta_{\rm cs}\gg1,
    \end{cases}
    \label{eq:enthalpy}
\end{equation}
where $\Theta_{\rm cs} \equiv T_{\rm cs}/m_e c^2$ denotes the dimensionless temperature in the current layer, and we used $\Gamma_{A}=5/3$ for $\Theta_{\rm cs}\ll 1$ and $\Gamma_{A}=4/3$ for $\Theta_{\rm cs} \gg 1$. Notice that for $\Theta_{\rm cs}\ll 1$, the long-wavelength regime $k\Delta_{\rm cs}\ll 1$ in Eq.~\ref{eq:KS_dispersion} requires expanding the Taylor series in Eq.~\ref{eq:enthalpy} to the next order in $\Theta_{\rm cs}$. In this case, the growth rate retains a dependence on $\Theta_{\rm cs}$.
Putting all the regimes together, we can describe the full parameter space in the asymptotic regime (long and short wavelengths w.r.t. $\Delta_{\rm cs}$) with the following dispersion relation:

\begin{equation}
    \frac{\omega^4}{k^2 g^2}\approx 
    \begin{array}{l}
    \left.
        \begin{cases}
        1,
            ~~~& k\Delta_{\rm cs}\gg1\\
        1,~~~& \Theta_{\rm cs}\ll k\Delta_{\rm cs}\ll 1\\
        \frac{k\Delta_{\rm cs}}{2\Theta_{\rm cs}},~~~& k\Delta_{\rm cs}\ll \Theta_{\rm cs}\\
        \end{cases}
        \right\}
        \Theta_{\rm cs}\ll 1;\\
        \left.
        \begin{cases}
        \frac{1}{9},
            ~&k\Delta_{\rm cs}\gg1\\
        \frac{1}{4}k\Delta_{\rm cs},
            ~&k\Delta_{\rm cs}\ll 1
        \end{cases}
        \right\}~~~
    \Theta_{\rm cs}\gg 1.\\
    \end{array}
    \label{eq:limit cases}
\end{equation}
The analytical growth rate for different values of $\Theta_{\rm cs}$ is shown in Fig.~\ref{fig:dispersion}, where the color of the line corresponds to the temperature in region {\bf A}. Asymptotic relations for the non-relativistic, $\Theta_{\rm cs}\ll 1$, and ultra-relativistic, $\Theta_{\rm cs}\gg 1$, regimes are shown with dotted black lines. In further sections, we employ the notation $\eta_g = \sqrt{g/\Delta_{\rm cs}}$ (ignoring $\tilde{h}$ and setting $k=1/\Delta_{\rm cs}$) as the fiducial KSI growth rate for a given thickness, $\Delta_{\rm cs}$, and gravity, $g$, to facilitate the comparison between different runs.

\begin{figure*}[htb!]
    \centering
    \includegraphics[width=\textwidth]{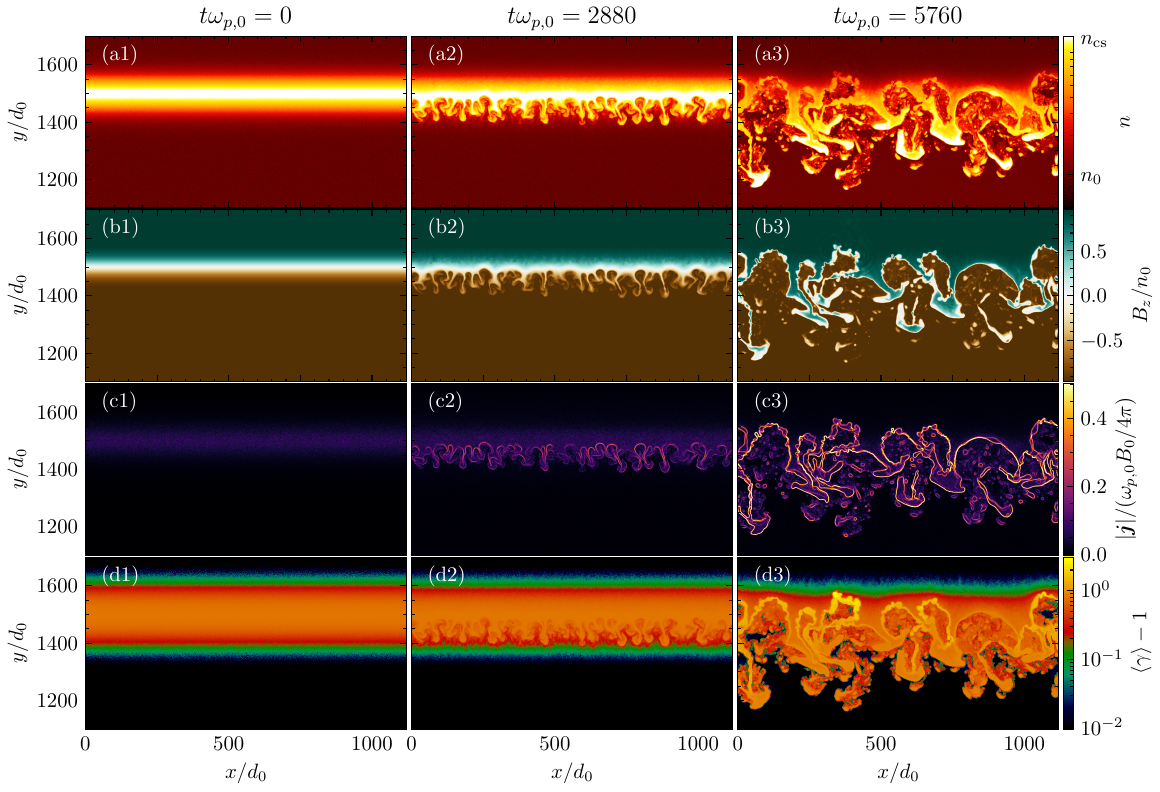}
    \caption{Time evolution of the KSI in a current sheet of initial thickness $2\Delta_0$ subjected to a force of strength $g_0$. Rows \textit{a}, \textit{b}, \textit{c}, and \textit{d} respectively display plasma number density, out-of-plane component of the magnetic field ($B_z$), current density ($|\bm{j}|$), and mean particle kinetic energy in units of the rest mass energy ($\langle\gamma\rangle - 1$). At late time (column 3), as regions with oppositely oriented fields come closer together, magnetic energy dissipation starts due to localized RDKI modes (panel \textit{c3}), energizing the plasma (yellow regions in panel \textit{d3}).}
    \label{fig:time_evol2d}
\end{figure*}

\section{Simulation results}
\label{sec:results}

\begin{figure*}[htb!]
    \centering
    \includegraphics[width=0.73\textwidth]{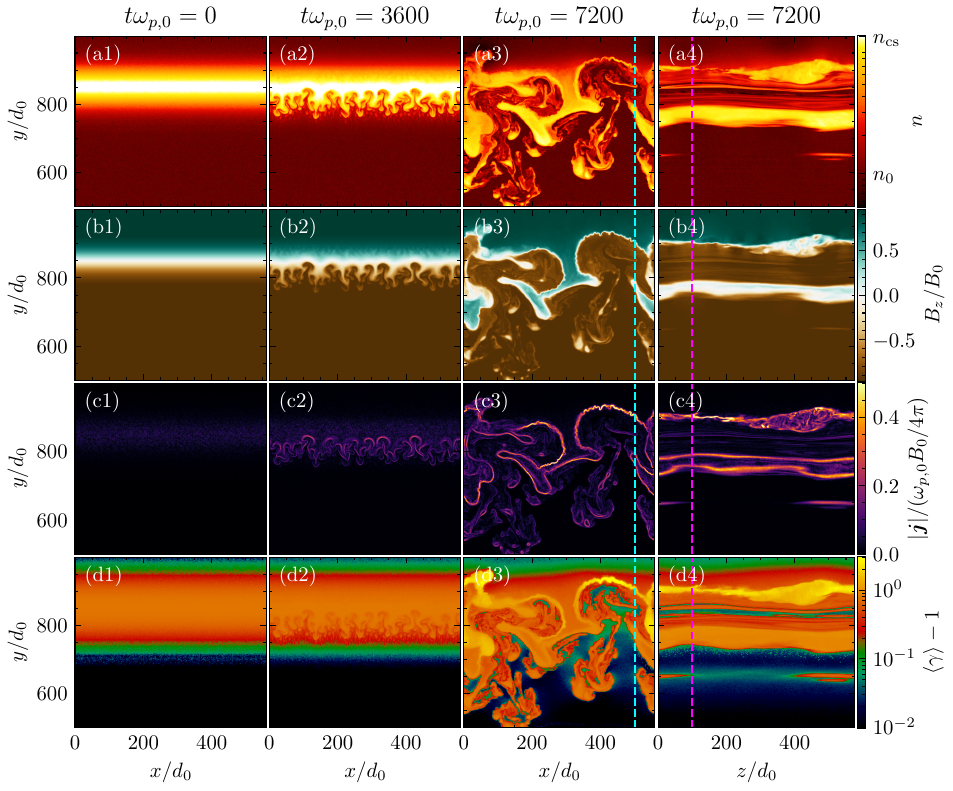}
    \caption{Snapshots of the same quantities as in Fig.~\ref{fig:time_evol2d} but for slices in the 3D run with $\Delta_{\rm cs}=2\Delta_0$ and $g=g_0$. The last column shows the slice in the $y$-$z$ plane where the upstream field lies. Magnetic reconnection is clearly visible in the $y$-$z$ plane, with the thin current layer at $z\approx 100\, d_0$ undergoing tearing instability, and a plasmoid ($z\approx 400\,d_0$) containing the energized particles. Dashed cyan and magenta lines indicate where the corresponding slices in $x$-$y$ and $y$-$z$ are taken.}
    \label{fig:time_evol3d}
\end{figure*}

\begin{figure}[htb!]
    \centering
    \includegraphics[width=\columnwidth, trim={1.7cm 9.5cm 7.2cm 6cm}, clip]{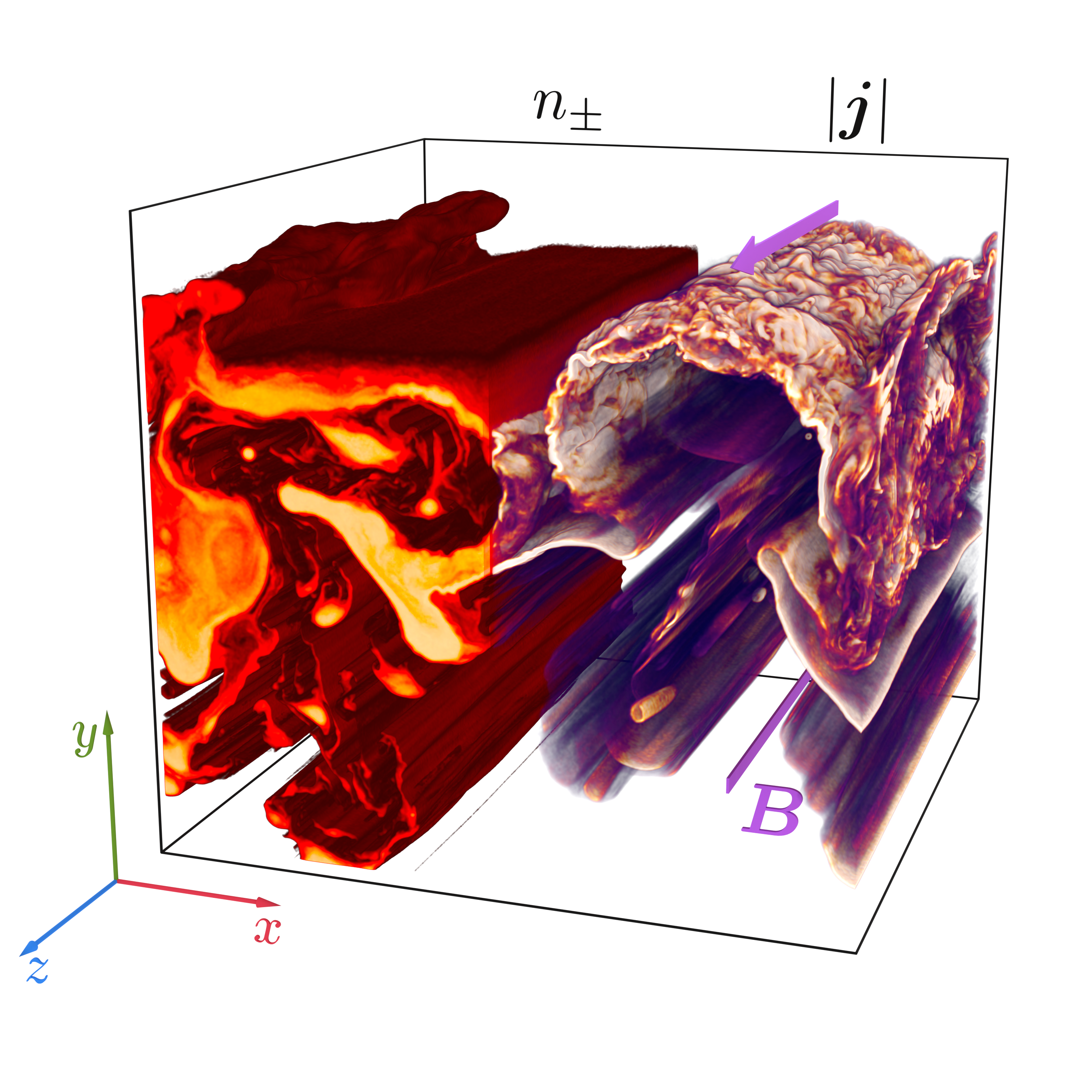}
    \caption{Volume rendering from the 3D simulation shown in Fig.~\ref{fig:time_evol3d} at $t\omega_{p,0}=7200$. We show on the left ($x<L_x/2$) the plasma density, while on the right ($x>L_x/2$) the electric current density.  
    Color scales are consistent with Fig.~\ref{fig:time_evol3d}. The direction of the upstream magnetic field is shown with arrows.}
    \label{fig:3dvol}
\end{figure}

In this section we present the results from both 2D and 3D simulations. In 2D, as the magnetic field initially is purely perpendicular to the plane of the simulation, the tearing instability cannot grow, and the only instability competing with the KSI is the relativistic drift-kink instability (RDKI; see also \citealt{zenitani_07,2021JPlPh..87f9013W}). As we demonstrate below, in both 2D and 3D, the initial dynamics of the layer is determined by the interplay of these instabilities. In 3D, we also observe the tearing instability in the plane of the magnetic field ($y$-$z$ plane) which grows during the non-linear stage of the KSI, when regions with opposing field polarities come together. In the following subsections, we study the evolution of the layer depending on various initial parameters for the 2D case, and also discuss the differences we observe between 2D and 3D.

\subsection{Reference case in 2D and 3D}
\label{sec:ref-case}


First, we focus on the evolution of our reference case in 2D with $g=g_0$, $\Delta = 2\Delta_{0}$, and $n_{\rm cs}/n_0 = 12$. Fig.~\ref{fig:time_evol2d} shows the current sheet at three different times. The left column shows the initial setup, while in the last column, $t\omega_{p,0}=5760$, the KSI is fully developed. As the KSI grows (at $t\sim 2000\, \omega_{p,0}^{-1}$), the hot plasma in the current sheet drips down into the magnetized region, creating non-linear fingers reminiscent of the Rayleigh-Taylor instability. The initial current layer in this case is wide enough, and the average current density is small enough, that the RDKI is not observed in the early stages. At later stages however, $t\gtrsim 5000\,\omega_{p,0}^{-1}$, as the current layer gets perturbed by the non-linear evolution of the KSI (see Fig.~\ref{fig:time_evol2d}, panel \textit{b3}), the local current density increases (see the bright regions in panel \textit{c3}), indicating that thinner, more intense current sheets are being formed. At this stage, small-scale ripples start to develop on the newly-formed thin current sheets due to the secondary RDKI (most notable near the top part of the layer at $x\approx 400\, d_0$, $y\approx 1500 \,d_0$ at $t=5760\,\omega_{p,0}^{-1}$). {The RDKI in the newly formed layers is then a ``parasitic'' instability, arising as a by-product of the non-linear evolution of the KSI.} The parasitic RDKI drives energy dissipation, which in turn energizes the plasma around the newly formed layers (panel \textit{d3}). The late-stage drift-kink-driven energy dissipation is similar to the findings by \cite{2021JPlPh..87f9013W}, who studied this instability in the context of an initially unstable current layer with no gravity. In Sec.~\ref{sec:energy-dissipation} we expand further on the energy dissipation history in our runs.

Fig.~\ref{fig:time_evol3d} shows the same analysis for a 3D box twice shorter in $x$ and $y$, and having $L_z \approx L_x$. The 3D run has the same physical parameters as in Fig.~\ref{fig:time_evol2d}: $g=g_0$, $\Delta = 2\Delta_{0}$, and $n_{\rm cs}/n_0 = 12$. Panels in the last column show a slice in the $y$-$z$ plane where the upstream magnetic field lies. The dynamics in the $x$-$y$ plane is very similar to our 2D simulation, with clear indications of RDKI developing in the late non-linear stages of KSI, when the interface between oppositely-directed fields becomes sufficiently thin. More importantly, in the $y$-$z$ plane the 3D run shows  that  at late stages the tearing instability starts to compete with the RDKI, and the sheet undergoes magnetic reconnection. This is clearly visible in panel \textit{c4} at $z\approx 100\,d_0$ and $y\approx 900 \,d_0$, where the current density is locally amplified. We also see a plasmoid being produced from this reconnection event (same panel, $z\approx 400\,d_0$), which contains hot plasma energized by the sites of active magnetic dissipation. 

\par We also show the plasma density and the current density at $t\omega_{p,0}= 7200$ from our 3D run in Fig.~\ref{fig:3dvol} with a volume rendering. In this figure, we show on the left ($x<L_x/2$) the plasma density, while on the right ($x>L_x/2$) the electric current density. The direction of the upstream magnetic field is shown with arrows. Regions of high overdensity (bright yellow on the left)---which track the hot plasma from the initial layer penetrating the strongly magnetized region---are surrounded by thin current-carrying layers (bright white on the right). In the upper part of the current density rendering (right side), one can see the interplay between tearing and drift-kink instabilities, which perturb the layer on small scales driving efficient energy dissipation.

\begin{figure*}[htb!]
    \centering
    \includegraphics[width=0.8\textwidth]{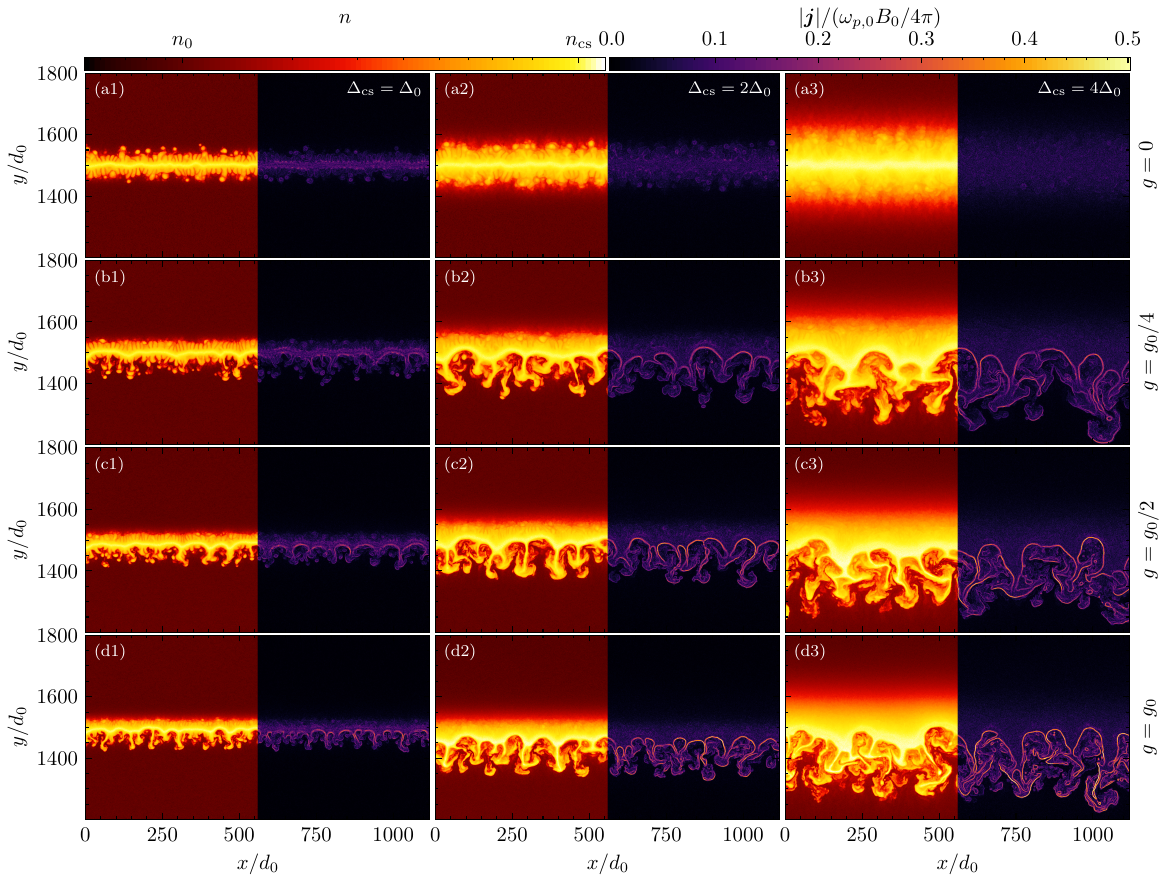}
    \caption{Snapshots of the plasma density and the electric current density at $ \eta_g\,t \approx 16$, where $\eta_g\equiv\sqrt{g/\Delta_{\rm cs}}$. Columns share the same current sheet widths of $\Delta_0$, $2\Delta_0$, and $4\Delta_0$ respectively, while rows share the same strength of gravity of $0$, $g_0/4$, $g_0/2$, and $g_0$ respectively. In the top row (panels \textit{a}), in which $g=0$, the snapshots were taken at the same time as in panels \textit{b} of the same column.}
    \label{fig:param_study2d}
\end{figure*}

\begin{figure*}[htb!]
    \centering
    \includegraphics[width=\textwidth]{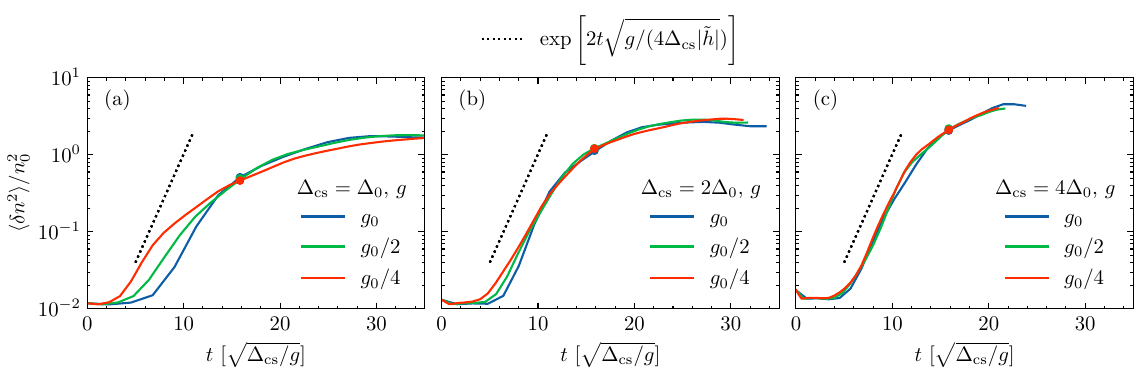}
    \caption{Time evolution of the box-averaged variance of the overdensity, $\delta n\equiv n - \langle n\rangle_x$, for the same simulations as in Fig.~\ref{fig:param_study2d}. Time is in units of $\sqrt{\Delta_{\rm cs}/g}$, as informed by  the fiducial growth rate. Panels \textit{a}, \textit{b}, and \textit{c} refer to different values of the initial thickness, while with color we represent different gravitational force strengths. The dashed line corresponds to the analytic growth rate for a mode with $k^{-1}=4\Delta_{\rm cs}$. Notice that all of these cases have the same value for the current layer temperature, and thus the same value for $\tilde{h}$. Circles indicate the times when the snapshots in Fig.~\ref{fig:param_study2d} were taken.}
    \label{fig:growth_rates}
\end{figure*}



\subsection{Dependence on $\Delta_{\rm cs}$ and $g$ in 2D}

To study how the KSI onset and evolution depend on physical parameters, we perform a series of 2D simulations varying $\Delta_{\rm cs}$ and $g$ ($n_{\rm cs}/n_0 = 12$ is fixed). In Fig.~\ref{fig:param_study2d}, we show snapshots of the plasma density (left), and the current density (right) near the current layer. The current density is normalized to $(c/4\pi)B_0/d_0$. Different rows correspond to different strengths of gravity, while different columns start with different widths of the layer. For easier comparison, the snapshots are taken at constant $\eta_g t$, where $\eta_g=\sqrt{g/\Delta_{\rm cs}}$.
For the $g=0$ case (upper row), we consider the same times as for $g=g_0/4$ (second row). In the fully non-linear regime, strong currents are induced at the boundaries of the KSI fingers, where two  regions of opposite magnetic polarity come close together. In cases when the initial current layer is thinner, $\Delta_{\rm cs} = \Delta_0$, thus the initial current is stronger (first column in Fig.~\ref{fig:param_study2d}), we observe the layer to corrugate before going KSI-unstable due to the faster-growing \textit{primary} drift-kink-instability (as opposed to the \textit{secondary} RDKI which occur in localized patches at later stages). In more realistic scenarios with wider initial layers (third column in Fig.~\ref{fig:param_study2d}), we see no evidence of the primary RDKI, and the dynamics is fully dictated by the linear and non-linear phases of the KSI. In all cases where gravity is present, the KSI dominates the dynamics below the layer. The upper half of the layer is relatively unperturbed, with only minor undulations present in cases with initially thinner layers (first column) due to the primary RDKI. In the non-linear stages of the KSI evolution, thin localized current layers are formed at the interface of KS fingers/bubbles. These localized layers are unstable to secondary RDKI modes, as we described above, and are ultimately responsible for magnetic energy dissipation (these layers are clearly highlighted in Fig.~\ref{fig:param_study2d} panel \textit{d3} along $y\approx 1500\, d_0$). 
As opposed to the primary RDKI, which is weaker for larger $\Delta_{\rm cs}$, the secondary RDKI is more prominent in cases with wider $\Delta_{\rm cs}$ (and stronger $g$). As anticipated above, the development of the secondary RDKI is parasitic, since it occurs in the wake of the late-stage evolution of the KSI.
In Appendix~\ref{sec:low-overdensity} we show that in cases where $n_{\rm cs}/n_0$ is smaller, a greater drift velocity is required in the initial layer to sustain the current, and thus we see much more pronounced evidence of  primary RDKI modes, which corrugate the layer before the onset of KSI, especially in cases where $\Delta_{\rm cs}$ is small.


To compare the growth rates of the KSI in our simulations with those predicted in Section~\ref{sec:analytic}, we calculate the variance in plasma number density following the definition by \cite{Gill2018}, $\delta n(x, y) = n(x,y) - \langle n\rangle_x$, where $\langle n \rangle_x \equiv \int n(x, y)dx / L_x $. We  measure the time evolution of the box-averaged $\langle\delta n^2\rangle / n_0^2$, which in the linear stage of the KSI should follow $\propto \exp{(2 \sqrt{kg/|\tilde{h}|}\, t)}$ for a mode with wavenumber $k$. Here, $|\tilde{h}|$ calculated from Eq.~\eqref{eq:enthalpy} using $\Gamma_{A}\approx 5/3$, which is appropriate for our case with relatively low temperature in the initial current sheet, $\Theta_{\rm cs}\approx 5/12$.
Fig.~\ref{fig:growth_rates} shows the box-averaged variance plotted vs time for all the cases presented in Fig.~\ref{fig:param_study2d}, where the dashed line corresponds to an exponential growth at the analytical growth rate for $k^{-1}\approx 4\Delta_{\rm cs}$ (a specifically chosen mode, which roughly corresponds to the dominant corrugation wavelength in Fig.~\ref{fig:param_study2d}). For all cases, time is measured in units of $\sqrt{\Delta_{\rm cs}/g}$. In all the cases (except for the case with smaller current sheet width $\Delta_{\rm cs}=\Delta_0$ and weaker gravity $g=g_0/4$) the linear stage growth rate matches well the analytic expectation. The case that deviates from the analytic prediction is also the one most affected by the initial corrugation of the layer due to the primary RDKI.

\begin{figure}[htb!]
    \centering
    \includegraphics[width=\columnwidth,clip]{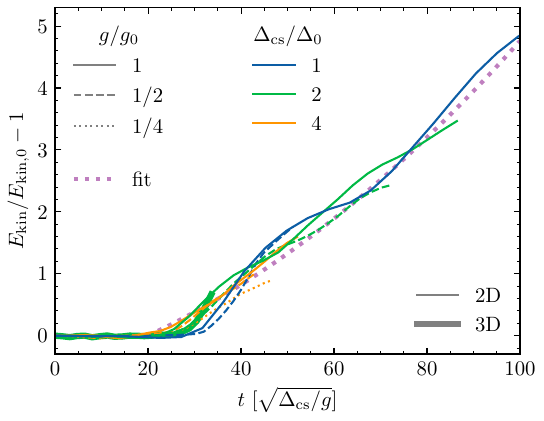}
    \caption{Estimate of magnetic energy dissipation, by  measuring the temporal evolution of the 
    {total kinetic energy of particles in the upper half of the domain, $E_{\rm kin}=\sum_i (\gamma_i-1)m_e c^2$}.
    We plot the fractional change $E_{\rm kin}/E_{\rm kin,0}-1= \Delta_{\rm eff}/\Delta_{\rm eff,0} -1$ as a function of 
    {$\sqrt{g/\Delta_{\rm cs}}\;t$.}
    Here, $\Delta_{\rm eff}$ is defined as the effective thickness of the dissipative region. The dotted purple line shows the solution of Eq.~\ref{eq:delta} with $\zeta=0.05$.}
    \label{fig:ke}
\end{figure}

\subsection{Magnetic energy dissipation}
\label{sec:energy-dissipation}
In the previous subsection we have demonstrated that the non-linear stages of the KSI lead to the formation of thin (skin-depth-thick) current layers, with lengths comparable to the dominant KSI wavelength. These thin layers go unstable to the RDKI and/or the tearing instability (the latter occurring only in 3D, for our geometry), which can {drive} efficient magnetic dissipation. 

The dissipated electromagnetic energy is converted into plasma energy. We quantify the dissipation efficiency in Fig.~\ref{fig:ke}, by measuring the temporal evolution of the 
{total kinetic energy of all particles in the upper half of the domain $E_{\rm kin}=\sum_i (\gamma_i-1)m_e c^2$.}
The figure employs a set of 2D simulations where we resolve the skin depth with $d_0=2.5$ cells and we adopt a larger box than in the fiducial runs described so far, in order to maximize the timespan covered by the simulations (see Section \ref{sec:setup} for details). The 3D run is shown with a thick green line.

When time is measured in units of $\sqrt{\Delta_{\rm cs}/g}$,  Fig.~\ref{fig:ke} shows that the fractional change $E_{\rm kin}/E_{\rm kin,0}-1$ follows the same temporal track regardless of $\Delta_{\rm cs}$ or $g$ (here, $E_{\rm kin,0}$ is the initial value). Within the limited timespan covered by our 3D simulation, we find that its curve overlaps with the corresponding 2D result, possibly indicating that the RDKI dominates magnetic energy dissipation even in 3D. We caution, however, that this conclusion might change when employing larger 3D domains (e.g., compare \citealt{zenitani_01} with  \citealt{sironi_spitkovsky_14}). The temporal evolution of $E_{\rm kin}$ is driven by the increase in the effective width $\Delta_{\rm eff}$ (along $y$) of the dissipation region. In fact, $E_{\rm kin}\propto \sigma_0 n_0 m_e c^2 \,L_x L_z \Delta_{\rm eff}\propto \Delta_{\rm eff}$, so the fractional change plotted on the vertical axis of Fig.~\ref{fig:ke} can be cast as $\Delta_{\rm eff}/\Delta_{\rm eff,0}-1$. More precisely, we define $\Delta_{\rm eff}=E_{\rm kin}/(\sigma n_0 m_e c^2 \,L_x L_z)$ and find that for a single layer, $\Delta_{\rm eff,0}=\Delta_{\rm cs}/(\Gamma_A-1)$, where $\Gamma_A$ is the adiabatic index in the hot layer; for $\Theta_{\rm cs}=(\sigma_0/2)(n_0/n_{\rm cs})=5/12$, we obtain $\Delta_{\rm eff,0}\simeq 2 \,\Delta_{\rm cs}$.
The effective width of the current layer should evolve as
\begin{equation}
\frac{d\Delta_{\rm eff}}{dt}=\zeta \frac{\Delta_{\rm eff}}{\tau_g}
\label{eq:delta}
\end{equation}
{with the characteristic timescale $\tau_g=\eta_g^{-1}\equiv \sqrt{\Delta_{\rm eff}/g}$}
as suggested by \citet{Lyubarsky2010}. The solution of this equation for $\zeta=0.05$ is overplotted in Fig.~\ref{fig:ke} with the dotted purple line, which provides a remarkably good fit to the simulation results. We therefore conclude that KSI-driven magnetic energy dissipation can be quantified by Eq.~\ref{eq:delta} with $\zeta=0.05$. This will be used in the following section to estimate the efficiency of KSI-driven dissipation in GRB and AGN jets.

\section{Conclusion and discussion}
\label{sec:conclusion}
We have studied magnetic energy dissipation in relativistic, accelerating striped jets. The effective gravity force $g=c^2 d\Gamma/dr$ in the rest frame of accelerating jets drives the KSI, which we have investigated by means of 2D and 3D particle-in-cell simulations. We find that the linear stage is well described by linear stability analysis, as derived by \citet{Lyubarsky2010} for relativistically hot layers and extended in this paper to the general case of arbitrary temperatures. The non-linear stages of the KSI generate thin (skin-depth-thick) current layers, with length comparable to the dominant KSI wavelength. There, the relativistic drift-kink mode (in both 2D and 3D) and the tearing mode (only in 3D, for our geometry) drive efficient magnetic dissipation. The dissipation rate can be cast as an increase in the effective width $\Delta_{\rm eff}$ of the dissipative, turbulent region, which follows $d\Delta_{\rm eff}/dt\simeq 0.05 \sqrt{\Delta_{\rm eff}\,g}$. Our (moderate-size) 3D simulation reveals the formation of reconnection plasmoids, yet the rate of field dissipation is roughly comparable to the corresponding 2D run. 

Our results have important implications for the location of the dissipation region in GRB and AGN jets, specifically as regard to the GRB ``prompt'' phase and the blazar-zone emission \citep{giannios_06b,giannios_12,mckinney_uzdensky_12, begue_17,giannios_uzdensky_19, gill_20}. In black-hole-powered striped jets, 
the typical stripe width $\ell$ (i.e., the distance between two consecutive current sheets) was estimated by \citet{giannios_uzdensky_19} to be $\ell\sim 1-100\, r_g$, where $r_g=G M/c^2$ is the gravitational radius of a black hole of mass $M$. Following \citet{Lyubarsky2010}, the Poynting flux of the jet is dissipated completely, $\Delta_{\rm eff}\sim \ell$, at 
\begin{equation}
r_{\rm diss}=12 \left(\frac{\Gamma_{\rm max}}{\zeta}\right)^2 \ell,
\end{equation}
where $\Gamma_{\rm max}$ is the Lorentz factor achieved if the Poynting flux is completely transformed into the plasma kinetic energy. Observations suggest that $\Gamma_{\rm max}\sim 10$ in AGN jets and $\Gamma_{\rm max}\sim 300$ in GRB jets. It follows that the expected dissipation distance in AGN jets is 
\begin{eqnarray*}
r_{\rm diss}&\sim &  1.2\cdot 10^5\,\left(\frac{\Gamma_{\rm max}}{10}\right)^2 \left(\frac{\zeta}{0.1}\right)^{-2} \ell\\
&\simeq &5.5 \left(\frac{\Gamma_{\rm max}}{10}\right)^2\left(\frac{\zeta}{0.1}\right)^{-2}\left(\frac{\ell}{r_g}\right)\left(\frac{M}{10^9M_\odot}\right)\rm pc.
\end{eqnarray*}
In long GRBs, having $\Gamma_{\rm max}\sim 300$, the dissipation distance is expected to be
\begin{eqnarray*}
r_{\rm diss}&\sim & 1.1\cdot 10^8\,\left(\frac{\Gamma_{\rm max}}{300}\right)^2 \left(\frac{\zeta}{0.1}\right)^{-2} \ell\\
&\simeq & 1.4\cdot 10^{14} \,\left(\frac{\Gamma_{\rm max}}{300}\right)^2 \left(\frac{\zeta}{0.1}\right)^{-2} \left(\frac{\ell}{r_g}\right)\left(\frac{M}{10 M_\odot}\right)\rm cm.
\end{eqnarray*}
While our simulations provide a framework for understanding KSI-driven magnetic dissipation in PIC for the first time, the predicted dissipation distance appears to be larger than what is inferred from observations of GRB and AGN jets. For instance, observations of the GRB prompt emission and blazar-zone emission suggest that efficient dissipation must occur at radii smaller than those predicted above \citep[e.g.,][]{ghisellini_10,giannios_spruit_07,giannios_08}. This discrepancy highlights the need for further investigation into factors that might enhance the efficiency of KSI-driven dissipation. One possibility is that the tearing mode, which is only present in 3D simulations, could play a significant role in accelerating the dissipation process. Another possibility is that the presence of an intense radiation field originating outside of the jet core could exert an additional drag force on the jet via Compton scattering, effectively increasing the gravitational acceleration and leading to faster dissipation. A thorough investigation of these possibilities will be presented in a forthcoming work.

\section{Acknowledgments}
L.S. acknowledges support from DoE Early Career Award DE-SC0023015, NASA ATP 80NSSC24K1238, NASA ATP 80NSSC24K1826, and NSF AST-2307202. This work was supported by a grant from the Simons Foundation (MP-SCMPS-00001470) to L.S. and facilitated by Multimessenger Plasma Physics Center (MPPC), grant PHY-2206609 to L.S. 

\bibliography{main.bib,araa.bib}
\bibliographystyle{aasjournal}

\appendix 
\section{Development of RDKI in layers with smaller overdensity}
\label{sec:low-overdensity}

In Sec. \ref{sec:ref-case}, we discussed that in the early stages of the simulation, layers with an initially stronger current density are more prone to RDKI (which we referred to as the \textit{primary} RDKI). We can understand this by looking at the RDKI growth rate, $\eta_{\rm RDKI}\lesssim \beta_D (c/\Delta_{\rm cs})$ \citep{zenitani_07}. Here, $\beta_D\ll 1$ is the drift velocity of pairs providing the electric current in the layer. For a fixed strength of the magnetic field and background plasma density, $\beta_D\propto (n_{\rm cs}\Delta_{\rm cs})^{-1}$, meaning that the RDKI growth rate scales as $\eta_{\rm RDKI}\propto n_{\rm cs}^{-1}\Delta_{\rm cs}^{-2}$. The KSI growth rate, on the other hand, scales as $\eta_g\approx \sqrt{g/\Delta_{\rm cs}}$. So for a fixed point in time in units of $\eta_g^{-1}$, which is what we used in Fig.~\ref{fig:param_study2d}, wider sheets are at earlier stages of the RDKI, which is why we only see significant RDKI in the left panels (\textit{a1}...\textit{d1}), where the sheet thickness is the smallest, $\Delta_{\rm cs}=\Delta_0$. 

\begin{figure*}[htb!]
    \centering
    \includegraphics[width=0.8\textwidth]{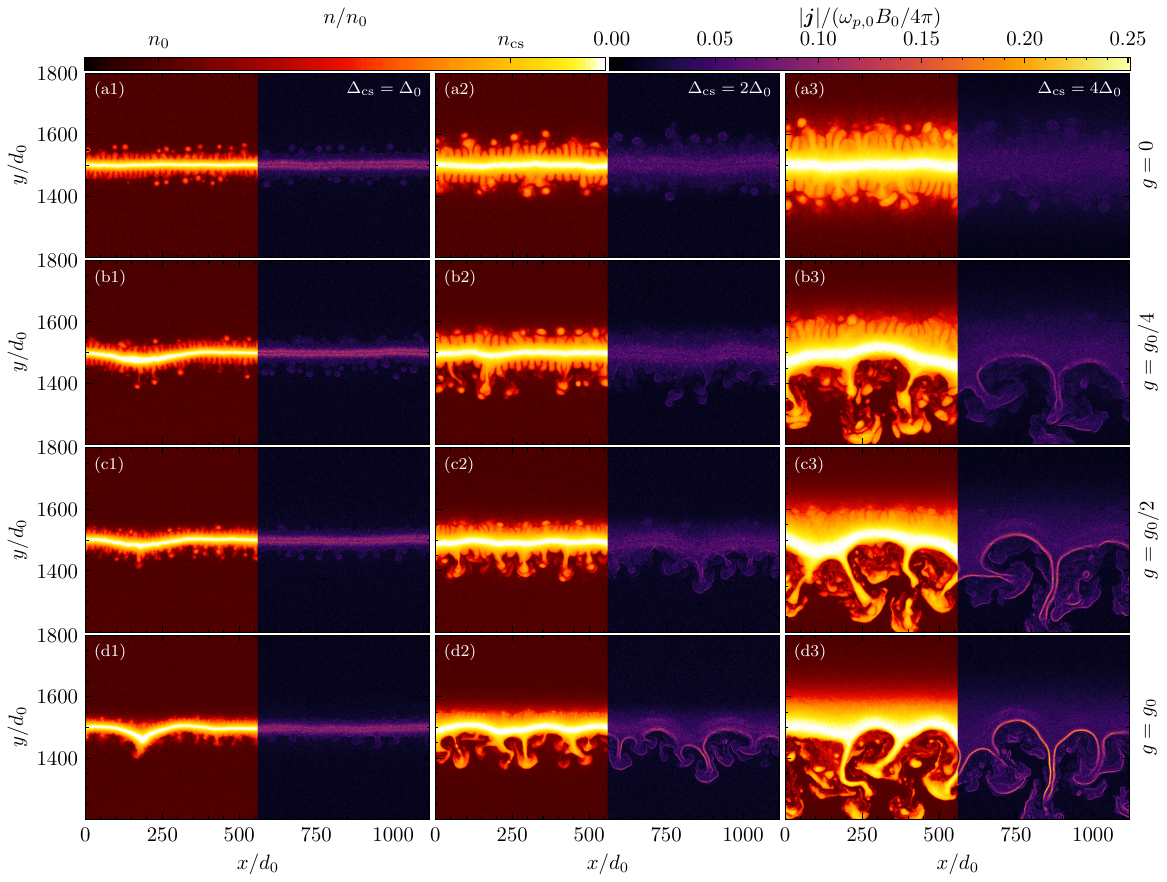}
    \caption{Snapshots of the plasma density and the electric current density at $\eta_g\,t \approx 32$, where $\eta_g\equiv\sqrt{g/\Delta_{\rm cs}}$, for simulations with overdensity of $n_{\rm cs}/n_0=3$. All panels are shown at twice the time of the corresponding panel from Fig.~\ref{fig:param_study2d} and refer to the same current sheet widths and gravity strengths as in Fig.~\ref{fig:param_study2d}. The KSI develops much slower in these cases as compared to colder cases.}
    \label{fig:param_study2d_od3}
\end{figure*}

In Fig.~\ref{fig:param_study2d_od3}, we show a series of simulations similar to those in Fig.~\ref{fig:param_study2d}, but for $n_{\rm cs}/n_0 = 3$. Snapshots are taken at exactly twice the times (in units of $\eta_g^{-1}$; twice, because of the temperature dependence of the growth rate) as in the original figure where the overdensity was $n_{\rm cs}/n_0 = 12$. Smaller overdensity also means faster growing RDKI; at times the snapshots in Fig.~\ref{fig:param_study2d_od3} are taken, we expect RDKI to be at later development stage than in Fig.~\ref{fig:param_study2d}. Indeed, in Fig.~\ref{fig:param_study2d_od3} we see the primary RDKI modifying the layer for all current sheet thicknesses. Importantly, for more realistic astrophysical parameters, where the width of the layer is macroscopic, i.e., $\Delta_{\rm cs}\gg d_0$, we expect the primary RDKI to play a sub-dominant role, which justifies why in the main text we have focused on cases with larger overdensity, where the RDKI is unimportant.



\end{document}